# Regularization of dielectric tensor tomography using total variation

HERVE HUGONNET,[1,2 †] SEUNGWOO SHIN,[3 †] AND YONGKEUN PARK[1,2,4,*]

[1]*Department of Physics, Korea Advanced Institute of Science and Technology (KAIST), Daejeon 34141, South Korea*
[2]*KAIST Institute for Health Science and Technology, KAIST, Daejeon 34141, South Korea*
[3]*Department of Physics, University of California at Santa Barbara, Santa Barbara, CA 93106, USA*
[4]*Tomocube Inc., Daejeon 34109, South Korea*
[†]*These authors equally contributed to the work.*
[*]*yk.park@kaist.ac.kr*

**Abstract:** Dielectric tensor tomography reconstructs the three-dimensional dielectric tensors of microscopic objects and provides information about the crystalline structure orientations and principal refractive indices. Because dielectric tensor tomography is based on transmission measurement, it suffers from the missing cone problem, which causes poor axial resolution, underestimation of the refractive index, and halo artifacts. In this study, we present the generalization of total variation regularization to three-dimensional tensor distributions. In particular, demonstrate the reduction of artifacts when applied to dielectric tensor tomography.

## 1. Introduction

Dielectric tensor tomography (DTT) is a recent development in microscopy that enables the reconstruction of the three-dimensional (3D) dielectric tensor distribution of a sample [1]. The dielectric tensor is a measurement of the 3D optical anisotropy, which fundamentally describes the light-matter interaction considering polarization of light, optical anisotropy, and molecular orientations. The intrinsic information of optically anisotropic materials, including the principal refractive indices (RIs) and crystalline structure orientation, can be obtained by diagonalizing the dielectric tensor. The unique ability of DTT to directly reconstruct the dielectric tensor is expected to be utilized in many disciplines, from metrology and soft-matter physics to biology [2-8]. Previous methods such as polarized light microscopy [9] (Fig. 1a), polarization-dependent digital holographic microscopy [10-14], polarization-dependent optical diffraction tomography [15-17] or fluorescence-based imaging techniques [18] can only provide limited information about the sample anisotropy or require some assumptions to retrieve the dielectric tensor. On the other hand, DTT can directly access the full dielectric tensor by solving a vectorial wave equation. However, DTT is based on light transmission three-dimensional quantitative phase imaging (QPI) measurements [1, 19], which makes it inherently affected by the missing-cone problem. Because of the limited numerical aperture of the used imaging system, some spatial frequencies of a sample cannot be reconstructed, resulting in poor axial resolution, low precision of reconstructed 3D orientations, underestimation of principal refractive indices, and halo artifacts [20].

Prior knowledge of a sample can be used to alleviate some of these artifacts. In many imaging systems, an image is known to have a non-negative value, which can be applied using the Gerchberg–Saxton algorithm [21, 22] or in combination with a denoising algorithm [23]. However, in DTT, the off-diagonal components of the dielectric tensor are often negative, depending on the orientations of the crystalline structures, hindering the use of such methods. Other minimization-based techniques, such as total variation (TV) regularization [24, 25] and other optimization methods using the image gradient [26-30] instead suppress missing cone artifacts by assuming the sparse presence of edges in the data. These methods are promising for the regularization of DTT data; however, their applicability to tensor data remains to be demonstrated. The last class of regularization uses deep learning to fill in missing data [31].

Here we propose and demonstrate the generalization of TV regularization to 3D tensor distributions. In particular, we focus on applying TV to DTT measurements, presenting that component-wise regularization of tensor data from DTT measurements. We show that using TV can greatly reduce missing cone artifacts: improving both principal RI and crystalline structure direction information.

## 2. Methods

### 2.1 Dielectric tensor tomography

Birefringence is a property of optical materials that causes light to diffract differently depending on the polarization. Birefringence resulting from supramolecular assemblies or crystalline structures can be physically described by a dielectric tensor [32]. The dielectric tensor is a generalized physical quantity from the dielectric constant that expresses the speed of light in a material but is dependent on the polarization. The dielectric tensor is symmetric and can be decomposed using singular value decomposition into three dielectric constants and three optical axes, with each dielectric constant corresponding to the speed of light for polarization parallel to a given optical axis [32].

As such, six unknown values must be obtained to completely determine the dielectric tensor. However, most optical systems can only change the illumination and detection polarizations, leading to four independent measurements that are insufficient for fully reconstructing the dielectric tensor. Using redundant spatial frequency information from slightly tilted illumination angles, DTT can reconstruct a full dielectric tensor [1, 33].

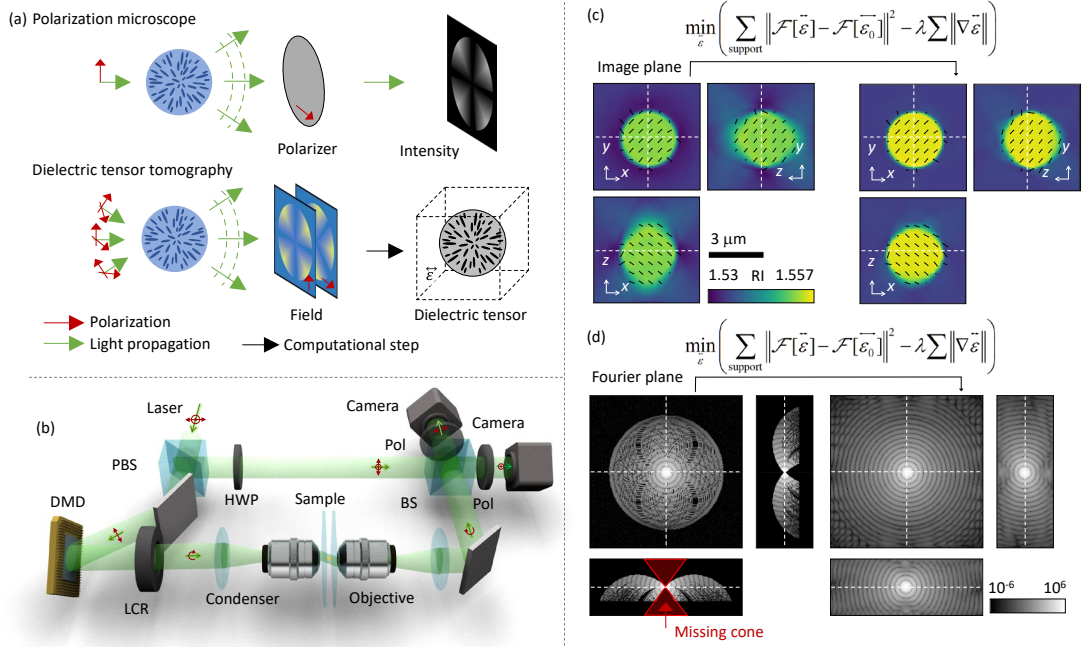

Fig. 1. (a) Schematic comparison between dielectric tensor tomography and polarization microscopy. (b) Optical setup, HWP: half-wave plate, LCR: liquid crystal rotator, DMD: digital micromirror device, PBS: polarized beam splitter and BS: beam splitter. (c) Effects of TV regularization in the image plane (d) Effects of the regularization in the Fourier plane.

DTT reconstruction is based on polarization-dependent light-field measurements obtained using off-axis holographic microscopy (Fig. 1b). For a given illumination wave vector $\vec{k}_{ill}$ and polarization $\vec{p}_i$, the field $\vec{\psi}_i$ transmitted through the sample is recorded. The illumination was then slightly tilted using $\Delta\vec{k}_{ill}$ to obtain a third field. Using three fields, the dielectric tensor $\ddot{\varepsilon}$ can then be expressed as

$$\ddot{\varepsilon}(\vec{r}) = \begin{pmatrix} \mathcal{F}^{-1}\left[-2ik_z \vec{p}_1 \cdot \mathcal{F}[\vec{\psi}_1](\vec{k}+\vec{k}_{ill})\right] \\ \mathcal{F}^{-1}\left[-2ik_z \vec{p}_2 \cdot \mathcal{F}[\vec{\psi}_2](\vec{k}+\vec{k}_{ill})\right] \\ \mathcal{F}^{-1}\left[-2ik_z \vec{p}_3 \cdot \mathcal{F}[\vec{\psi}_3](\vec{k}+\vec{k}_{ill})\right] \end{pmatrix} \begin{pmatrix} \vec{p}_1 \\ \vec{p}_2 \\ (1-i\vec{r} \cdot \Delta\vec{k}_{ill})\vec{p}_3 \end{pmatrix}^{-1}.$$

Using $\vec{k} = \begin{pmatrix} k_x & k_y & k_z \end{pmatrix}^T$, the Fourier space coordinate system. This process is repeated for several illumination angles to populate the Fourier space (Fig. 1d) [1, 34].

*2.1 Total variation regularization for dielectric tensor tomography*
In DTT, although the full dielectric tensor can be retrieved for accessible spatial frequencies, not all frequencies are accessible. Because DTT is based on transmitted light in an optical setup with a limited numerical aperture, it suffers from a missing cone problem [35]. The name of the missing cone problem comes from the fact that non-accessible spatial frequencies form a cone in the Fourier space (Figs. 1c–d). The missing cone problem is a common problem in tomographic measurements with a limited illumination angle [20, 36] and also occurs in X-ray computed tomography [37], three-dimensional transmitted electron microscopy [38], optical diffraction tomography [39-42], and widefield fluorescence imaging [43, 44].

Various computational methods, called regularization, have been developed to fill in the missing information. A known constraint of a sample was used to predict inaccessible spatial frequency information. One constraint that can be used is edge sparsity, in which the image is supposed to have a limited number of edges. This constraint can be applied using a total variation (TV) algorithm [45].

The TV algorithm minimizes the average gradient norm while preserving the fidelity of measured data. This can be mathematically expressed as follows:

$$\min_{\vec{\varepsilon}} \left( \sum_{\text{support}} \left\| \mathcal{F}[\vec{\varepsilon}] - \mathcal{F}[\vec{\varepsilon}_0] \right\|^2 - \lambda \sum \left\| \nabla \vec{\varepsilon} \right\| \right),$$

where $\vec{\varepsilon}_0$ is the retrieved dielectric tensor using the DTT algorithm, $\vec{\varepsilon}$ is the regularized dielectric tensor, "support" denotes accessible spatial frequencies, $\lambda$ is the regularization constant, and $\nabla$ is the gradient applied independently to each element of the tensor. Minimization is then conducted using a gradient-based method, that is, the FISTA method [24].

We further improve the existing TV algorithm by using a multiscale approach to speed up the computation time. This is particularly important for DTT because of the large size of the tensorial data compared to scalar imaging. We experimentally noticed that TV algorithms are often very fast at removing high-frequency noise, but the suppression of low-frequency missing cone artifacts often requires several iterations. For this reason, we suppressed missing cone artifacts using an under-sampled low-resolution version of the data by using several iterations and then suppressed high-frequency artifacts at full resolution by using only a few iterations. Owing to the lower execution time when using low-resolution data, this strategy reduces the overall computation time while preserving imaging quality.

For the simulation, we chose to use a small regularization constant to only fill up data in the missing cone region, but we used a larger regularization constant for the experiment because it enables faster convergence rates and has a denoising effect on experimental data [24]. We used $\lambda = 5 \cdot 10^{-4}$ for simulations and $\lambda = 10^{-2}$ for experiments; at low resolution, a total of 200 inner and outer iterations were performed, whereas at high resolution, 100 inner iterations and 20 outer iterations were used. The execution time of the regularization algorithm for $218 \times 218 \times 80 \times 3 \times 3$ data on a workstation computer with a graphics processing unit (GTX1070ti, Nvidia) was 69 s.

### 2.3 Optical setup

The setup used to demonstrate the proposed method experimentally is shown in Fig. 1b. Based on a Mach-Zehnder interferometer, the laser beam was first split into a reference beam and a sample beam. A digital micromirror device was used to systematically control the illumination angle of a plane wave [46, 47], followed by a liquid crystal retarder to control the polarization of the beam. The sample beam was then projected onto a sample using a water immersion condenser (UPLSAPO 60XW, NA=1.2, Olympus). Then, the scattered light was collected using an oil immersion objective (PLAPON 60X, NA=1.42). Finally, the sample beam interfered with the reference beam on two cameras (Lt425R, Lumenera) with an orthogonal analyzer to obtain polarization-dependent fields.

## 3. Results

### 3.2 Quantification of reconstruction quality improvement

To evaluate the efficiency of TV regularization in reducing missing cone artifacts, we first simulated light scattering in a synthetic liquid crystal distribution. In particular, we chose hybrid aligned nematic (HAN) and twisted nematic (TN) distributions because of their importance in liquid-crystal display technology [48, 49].

Simulations were carried out on a 5-μm-height and 5-μm-diameter cylinder containing the aforementioned distributions. The transmitter field was computed using the finite-difference time-domain method (Lumerical FDTD 2020), and tomograms were obtained using the DTT algorithm followed by TV regularization. The fidelity was measured against the original phantom using three criteria: the orientation angle, axial resolution, and RI value. Reconstructed tomograms are shown in Fig. 2a for visual inspection.

First, we examined the crystalline orientation. In the HAN distribution, the altitude angle varies linearly from 0° to 90°, whereas in the TN distribution, the liquid crystal is rotated along the azimuthal angle. Figures 2bi–ci show the average angle of the crystalline orientation as a function of the vertical position. The expected, non-regularized, and TV-regularized distributions were plotted. TV regularization improves the orientation fidelity for both distributions. This improvement can be evaluated using the standard deviation of the expected distribution. The HAN standard deviation improved from 13.9° to 13.6°, whereas the TN standard deviation improved from 15.9° to 6.5°.

TV regularization also improves the resolution and RI quantification. Figures 2bii-cii show the RI distribution at the center of the cylinder as a function of the vertical position. Again, the use of TV regularization corrects RI underestimation and improves the resolution, as seen from the sharp increase in the RI near the cylinder boundary.

The TN distribution showed great fidelity after TV regularization. However, the HAN distribution, while showing improvement when using TV regularization, still suffers from a significant RI underestimation. We believe this to be due to the liquid crystal alignment parallel to the optical axis, which reduces the SNR and makes the system more sensitive to other issues, such as multiple scattering.

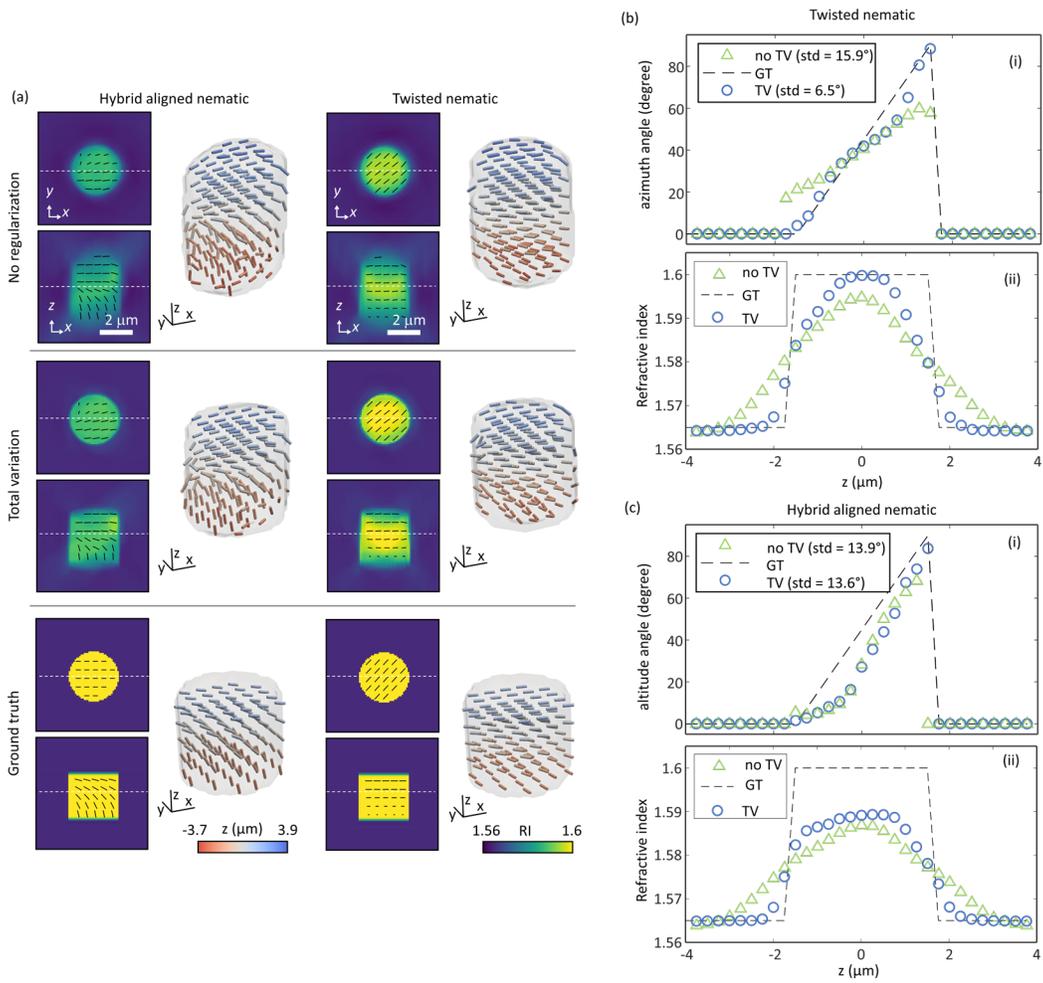

Fig. 2. (a) 3D visualization of the different liquid crystal distribution (b) plot of the azimuth angle (i) and RI (ii) for the twisted nematic distribution as a function of depth (c) plot of the altitude angle (i) and RI (ii) for the hybrid aligned nematic distribution as a function of depth

## 3.3 Liquid crystal droplet sample

While simulations are important to quantify the fidelity improvement after regularization, their applicability to experimental data also needs to be demonstrated. In this study, we set out to image liquid crystal droplets. Both radial and bipolar liquid crystal droplets were prepared following the protocol in [1] and were subsequently imaged. The reconstructed results showed the expected crystalline structure. Furthermore, when using TV regularization, missing cone artifacts are minimized. Indeed, the principal RI increased, hallo artifacts decreased, and vertical cross-sections appeared sharper (Fig. 3).

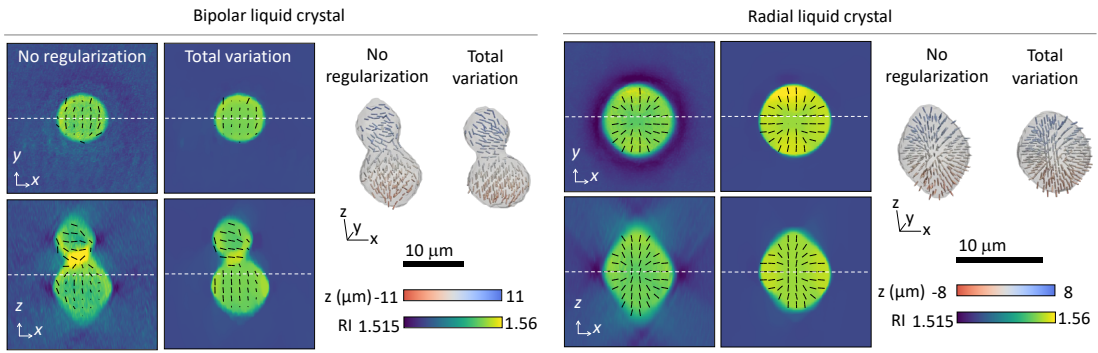

Fig. 3. TV regularization effect on experimental data.

## 4. Discussion

We presented that TV regularization was able to reduce missing cone artifacts in both the simulation and experimental DTT data. Using both the numerical and experimental studies, we demonstrated that the present approach could significantly enhance the reconstruction accuracy and fidelity of DTT by filling up the information in the missing cone by exploiting the regularization. The crystalline direction fidelity, RI underestimation, and axial resolution were improved. Numerical simulations of the HAN and TN liquid crystals showed the highly precise retrieval of dielectric tensor values. Experimental results also showed significantly enhanced axial resolution, signal-to-noise ratio, and also clear tomographic visualization of crystalline directions. While the applicability of TV regularization has been demonstrated on DTT data, the algorithms and concepts presented here can be extended to other tensor imaging modalities, such as diffusion tensor imaging [50, 51] or anisotropic Brillouin microscopy [52].

Although the present method requires additional processing time compared to applying the regularization to DTT, several approaches may be used to address this speed issue. For example, high-performance graphic processing units can be employed with parallel computations. Also, deep learning approaches can be used to apply the TV algorithm to avoid time-consuming iteration steps [53, 54]. Then, the present method can be further extended to the time-lapse quantitative analyses of various phenomena related to optical anisotropy.

Another limitation of the present method is the presence of speckle noise, resulting from a highly temporally coherent light source and an off-axis Mach-Zehnder interferometric microscopy. Recently, it was shown that a low-coherence light source could be used for reconstructing three-dimensional reconstruction of RIs of a sample by carefully matching the optical path lengths in interferometry [55], deconvolution microscopy with partially coherent light [56], Fourier ptychographic diffraction tomography [57], or space-domain Kramers–Kronig relations [58, 59]. When such a tomographic approach was used in terms of a polarization-sensitive manner, it can be used to reconstruct DTT [60]. Then, the current approach can be readily applied to the DTT reconstructed with a low-coherence light, which may significantly reduce speckle noise and thus improve signal-to-noise ratios.

In view of the unique label-free and quantitative imaging contrast capability exhibited by dielectric tensors down to diffraction-limited resolutions, it could potentially be used for the quantitative imaging and analysis of topological defects in liquid-crystal materials [61]. Moreover, the dielectric tensor information has considerable advantages in histopathological applications over standard polarization microscopy or staining methods, particularly in assay time and cost, and it can also provide quantitative tomographic evaluations of collagen fiber structures which is related to tumor metastasis [61]. Going forward, we envision that this synergistic approach of regularization and DTT could have broad applications, possibly in conjunction with newly emerging active soft materials [62].

**Funding.** This work was supported by the KAIST UP program, BK21+ program, Tomocube Inc., National Research Foundation of Korea (2015R1A3A2066550), KAIST Institute of Technology Value Creation, Industry Liaison Center (G-CORE Project) grant funded by the Ministry of Science and ICT (N11210014, N11220131), and Institute of Information & Communications Technology Planning & Evaluation (IITP; 2021-0-00745) grant funded by the Korean government (MSIT).

**Disclosures.** The authors declare no conflicts of interest.

**Data availability.** Data underlying the results presented in this paper are not publicly available at this time but may be obtained from the authors upon reasonable request.